\documentclass[doublecol]{epl2} 
\usepackage{amsmath}
\usepackage{graphicx}
\usepackage{xcolor}
\usepackage[pdfstartview=FitH,colorlinks=true,citecolor=blue,linkcolor=blue,urlcolor=blue]{hyperref}
\usepackage{ulem}

\title{Stationary transport above the critical velocity in a one-dimensional superflow past an obstacle}

\shorttitle{Stationary transport above the critical velocity for superfluidity}  

\author{J. Huynh\inst{1} \and F. H\'ebert\inst{1} \and P.-É. Larré\inst{1} \and M. Albert\inst{1,2}}

\shortauthor{J. Huynh \etal}

\institute{                    
  \inst{1} Universit\'e C\^ote d'Azur, CNRS, Institut de Physique de Nice, 06200 Nice, France\\
  \inst{2} Institut Universitaire de France (IUF)
}

\abstract{
We consider in this work the different possible stationary flows of a one-dimensional quantum fluid in the mean-field regime. We focus on the supersonic regime where a transition from a time-dependent flow to a stationary diffractive flow occurs at a given critical velocity. We give nonperturbative results for this critical velocity in the presence of a localised obstacle of arbitrary size and strength. In addition, we discuss the existence of superfluid-like solution in the supersonic regime due to resonant transport and provide a complete map of the different regimes of stationary transport of a quantum fluid.
}

\begin{document}

\maketitle

\begin{figure}[!t]
\centering
\includegraphics[width=.94\linewidth]{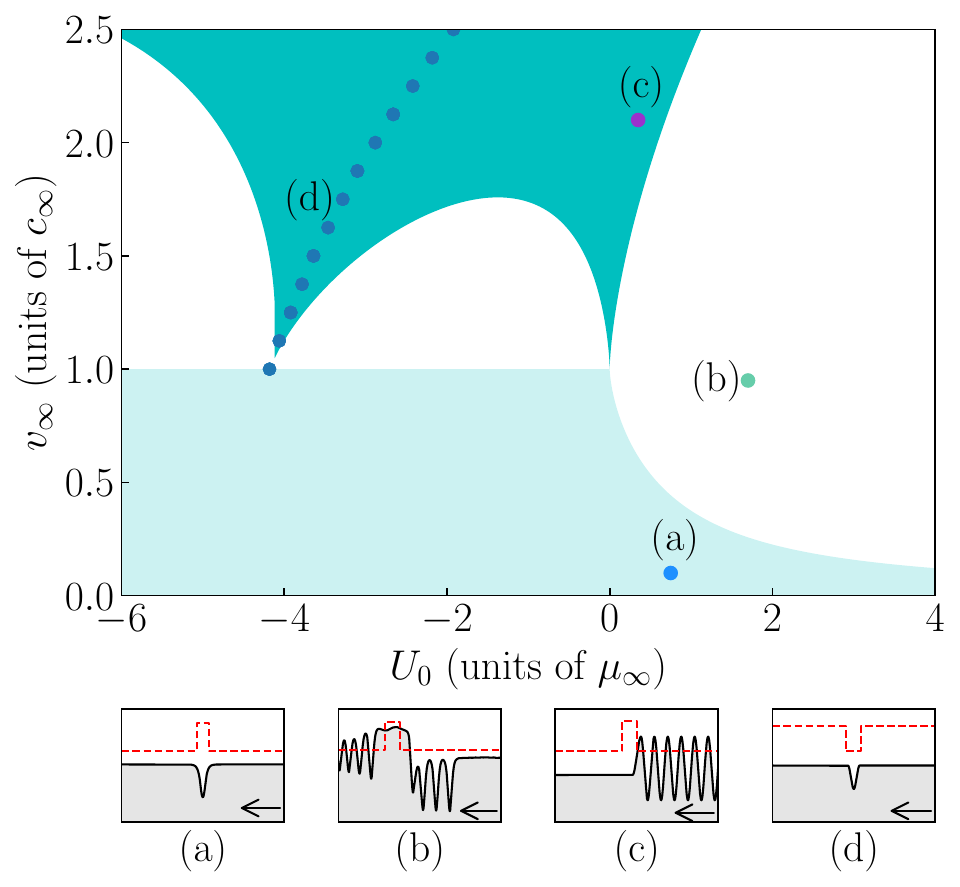}
\caption{\label{fig:phase} Typical phase diagram of the possible stationary flows in the presence of a localised obstacle, hereby a square potential, as a function of the fluid velocity $v_\infty$ and the strength of the obstacle $U_0$. The different regimes range from a superfluid (light blue) to a normal stationary regime  (dark blue), the white phase in between corresponding to the nonstationary nonsuperfluid regime. The dotted curve corresponds to a resonant state where the supersonic solution mimics the superfluid ones. Generic space dependent density profiles $n(x)$ are given in subfigures (a), (b), (c) and (d), for the corresponding points in the phase diagram. This phase diagram was obtained for a cubic nonlinearity $g(n)=n$ and a square potential of width
$\sigma=1$ and amplitude $U_0$. The units are detailed in the text.}
\end{figure}

\section{Introduction} \label{sec:Introduction}
 
One important property of superfluids is their ability to move without dissipation below a certain critical velocity $v_c$ \cite{Leggett1999}. A phase transition occurs at $v_c$ and the superfluid behaves as a normal dissipative fluid for velocities larger than $v_c$. First observed in liquid Helium \cite{Kapitza1938,Allen1938}, superfluidity was later shown to be more generic and was observed in various quantum fluids \cite{Osheroff1972,Raman1999,Amo2009,Michel2018}. Soon after its discovery, the critical velocity was theorised by Landau \cite{Landau1941a,Landau1941b} who proposed a very elegant and general criterion which states that $v_c=\min_{\boldsymbol{p}}\varepsilon (\boldsymbol{p})/p$ where $\varepsilon(\boldsymbol{p})$ is the spectrum of elementary excitations with momentum $\boldsymbol{p}$. However, this prediction usually overestimates the actual critical velocity and was verified experimentally only under very specific configurations as, for instance, by moving a single ion in liquid Helium \cite{McClintock1974}. The reason is that Landau's argument is perturbative and therefore does not take properly into account the nonlinear nature of the problem of interaction between quantum fluids and external potentials. Important progresses arose with the introduction of a simpler model to describe the flow of a quantum fluid: The nonlinear Schr\"odinger (NLS) equation or Gross-Pitaevskii (GP) equation \cite{Ginzburg1958,Gross1961,Pitaevskii1961}. In particular, nonperturbative results were derived for the first time by Frisch and collaborators in two dimensions in the presence of an impenetrable cylinder \cite{Frisch1992} followed by a series of works for various obstacles (see ref. \cite{Huynh2022} for a review). Although this model is not satisfactory for the description of dense systems such as liquid Helium, it is very accurate for weakly interacting superfluids such as Bose-Einstein condensates \cite{Pitaevskii2016} or quantum fluids of light \cite{Carusotto2013}.

However, the transport properties of quantum fluids described by a NLS equation are not restricted to superfluidity and display a rich phenomenology \cite{PavloffLeboeuf2001,Engels2007,Dries2010,Eloy2021} which is summarised in fig.~\ref{fig:phase}, for a one-dimensional system.
In the presence of a localised obstacle, three different regimes of transport exist depending on its strength and relative velocity with the fluid. Below the actual superfluid critical velocity $v_c$, the flow is stationary and only locally perturbed in the vicinity of the obstacle, as illustrated in subfigure (a) of fig.~\ref{fig:phase}. Above this threshold, which strongly depends on the obstacle, the flow can no longer be stationary due to the continuous emission of linear and nonlinear excitations (fig.~\ref{fig:phase} (b)) which leads to a slowdown of the superfluid motion and possibly to wave and quantum turbulence \cite{Nazarenko2011,Barenghi2014}. At larger velocities, a second critical velocity, often referred to as the supersonic separatrix \cite{PavloffLeboeuf2001, Leszczyszyn2009, Kamchatnov2012}, separates the latter regime from another regime of stationary transport. We will denote this second critical velocity $v_s$. The high-velocity stationary regime observed for velocities larger than $v_s$ is reminiscent of the linear Schr\"odinger equation since the kinetic energy becomes much larger than the interaction energy in the fluid. In that case, the flow is partly backscattered by the obstacle and generally experiences friction.
The incoming and reflected flows interfere and create a standing wave with a density modulation ahead of the obstacle (fig. \ref{fig:phase} (c)).
Yet some configurations exist in which dissipation does not occur, even in the nonsuperfluid phase, due to resonant transport \cite{Paul2005, Rapedius2006, Rapedius2008}. For specific obstacle parameters there might exist curves in the supersonic stationary phase where backscattering is suppressed and the fluid experiences no drag at all  mimicking a superfluid solution (fig.~\ref{fig:phase} (d)), a behaviour normally present below the superfluid critical velocity $v_c$, in the subsonic regime\cite{ParisMandoki2016}.

The aim of this letter is to determine, in a nonperturbative way, the supersonic separatrix – i.e. the border between the nonstationary and the stationary nonsuperfluid regimes – for a generic quantum fluid flowing past a simplified localised obstacle in the one-dimensional mean-field regime. In addition, we study in details the conditions to obtain superfluid-like solutions in the supersonic regime. Combined with previous results for the superfluid critical velocity \cite{Huynh2022}, this work provides a complete 
map of the different possible regimes of stationary transport for a one-dimensional quantum fluid, above and below the sound velocity, for repulsive or attractive obstacles, and for different types of nonlinearities.

This paper is divided as follows: The model of the quantum fluid, based on a generalisation of the 1D nonlinear Schrödinger equation to any local self-interaction potential increasing with the fluid density, is first detailed. This general approach makes it possible to describe many superfluid systems ranging from ultracold atomic Bose and Fermi gases \cite{Pitaevskii2016} to exciton-polariton condensates in semiconductor optical microcavities \cite{Amo2009,Carusotto2013} and fluids of light \cite{Michel2018,Fontaine2018,Vocke2016,Eloy2021,Leboeuf2010,Larre2015,Santic2018}. A thorough analytical study is then performed in the following sections in the limits of narrow or wide obstacles. Finally, we bridge the gap between these two limiting cases with a numerical study for a model obstacle and characterise analytically the perfect transmission lines.

\section{Theoretical model}
\label{sec:Model}
We consider a one-dimensional quantum fluid flowing in the negative-$x$ direction in the framework of the NLS equation. For the sake of clarity, we employ here the language of weakly interacting bosonic particles of mass $m$ although the results derived in this paper are of wider interest. A quantum fluid dictionary is provided in Supp. Mat. for readers interested in other physical realisations of this model. The dynamics of the considered system is governed by a generalised nonlinear Schrödinger equation for the order parameter $\psi$
\begin{equation}
\label{eq:GNLSE}
i\hbar\partial_{t}\psi=\left[-\frac{\hbar^{2}}{2m}\partial_{xx}+U(x)+g(|\psi|^{2})\right]\psi.
\end{equation}
The flow is here constrained by an obstacle described in eq.~\eqref{eq:GNLSE} by a potential $U(x)= U_0 f(|x|/\sigma)$ which attains its single positive maximum (negative minimum) $U_0$ at $x = 0$ and which is localised, i.e., which vanishes as $|x|\gg \sigma$, with $\sigma$ being its typical range. Throughout this work, we will exemplify our results with a repulsive (attractive) square potential $U(x)=U_{0}\Theta(\sigma/2-|x|)$ but results with a Gaussian potential are given in the Supp. Mat. The reason why we employ such a toy model is because it allows to obtain analytical results without loss of generality. In addition to the external potential, the fluid is also subjected to a self-interaction described by the local nonlinear term $g(|\psi|^{2}=n)\psi$, where the potential $g(n)$ is an increasing function of the density $n$. 
We consider two different kinds of nonlinearities. 
We mostly used the cubic nonlinearity of the standard NLS equation, where $g(n)=n$ in appropriately chosen units. This describes accurately dilute ultracold bosonic atoms \cite{Huynh2022}. We also considered a saturable nonlinearity of the form $g(n)=(1+n_s)^2 n / \left[n_s(n+n_s)\right]$ which describes quantum fluids of light in rescaled units, where $n$ is related to the light intensity and $n_s$ to the saturation intensity in the nonlinear medium (see Supp.~Mat.).

We now look for the existence of out-of-equilibrium stationary solutions of eq.~\eqref{eq:GNLSE} of the form
$  \psi(x,t)=\exp[-i\mu t/\hbar] A(x) \exp[i\varphi(x)]$,
from which the density and the velocity fields are obtained from $n(x)=A(x)^2$, $v(x)=\hbar \varphi'(x)/m$ and $\mu$ is the chemical potential. This yields the following equation of motion for these fields
\begin{eqnarray}
\label{eq:HydroEqs}
&n(x)v(x)=\Phi,& \\
&-\displaystyle{\frac{\hbar^2}{2m}}A''(x)+\left[U(x)+g(n)+\frac{m\Phi^2}{2A(x)^4}\right]A(x)=\mu A(x).&\nonumber
\end{eqnarray}

The first of eqs.~\eqref{eq:HydroEqs} is simply the current conservation while the second one expresses the space dependence of the density and therefore the velocity through current conservation. These equations have to be complemented with boundary conditions. As explained in ref. \cite{PavloffLeboeuf2001}, a regime of stationary flow exists for supersonic velocities, but in this case the radiation condition \cite{Lamb1997} requires that the wake is always located ahead of the obstacle, i.e. upstream, with no long-range perturbation of the fluid on the downstream region, where the flow remains unperturbed. The solution has therefore to tend to a constant solution with density $n_\infty$ and velocity $v_\infty$ far away from the obstacle in the downstream region (in our case $x\to -\infty$) with $\Phi=n_\infty v_\infty$ and $\mu=\frac{1}{2}mv_\infty^2+g(n_\infty)$. For comparison, a stationary superfluid solution satisfies the same condition in the upstream and downstream regions which is way more restrictive. Finally, two important scales emerge due to the nonlinearity $g(n)$, namely the sound velocity $c_\infty$ and the healing length $\xi_\infty$. They are defined in the downstream region as $mc^2_\infty=n_\infty g'(n_\infty)=\mu_\infty$ and $\xi_\infty=\hbar/mc_\infty$. In the rest of the manuscript we rescale all quantities in terms of $n_{\infty}$ for densities, $c_{\infty}$ for velocities, $\xi_{\infty}$ for distances, and $mc_{\infty}^{2}$ for energies. This corresponds to the substitution $\hbar=m=1$ and $\mu=g(1)+\frac{1}{2}v_\infty^2$ in eq.~\eqref{eq:HydroEqs}.

\begin{figure}[!t]
\centering
\includegraphics[width=.96\linewidth]{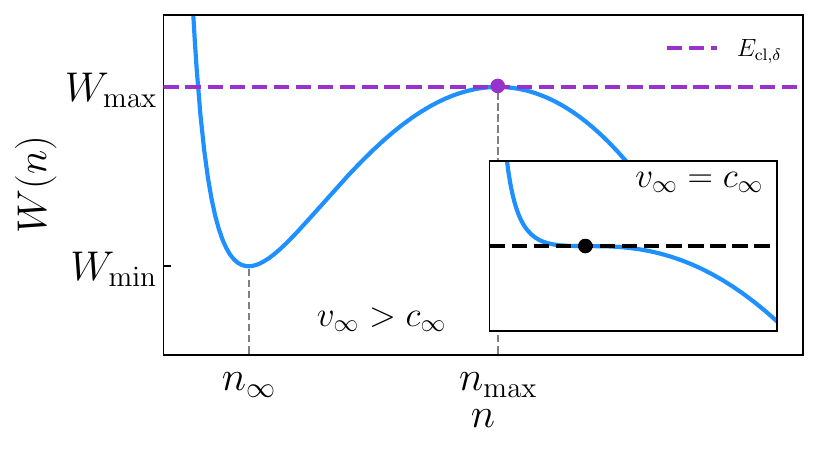}
\vspace{-0.4cm}
\caption{\label{fig:Fict_Pot}Schematic behaviour of the fictitious potential $W(A^2=n)$ for a cubic nonlinearity of the form $g(n)=n$ and $v_\infty>c_\infty$. The inset displays the special case $v_\infty=c_\infty$.}
\end{figure}

The main objective is now to search for the condition of existence for the solutions to eq.~\eqref{eq:HydroEqs}, which depends on the value of the injection velocity $v_{\infty}$. The last value under which there is no longer a solution to eq.~\eqref{eq:HydroEqs} defines the equation of the supersonic separatrix. In the spirit of ref. \cite{PavloffLeboeuf2001}, eq.~\eqref{eq:HydroEqs} can be rephrased in terms of a Hamilton equation describing the dynamics of a fictitious classical particle of position $A(x)$ and momentum $p=A'(x)$ at time $x$. The corresponding Hamilton function reads \cite{Paul2007}

\begin{equation}
\mathcal H(A,p) = \frac{p^{2}}{2} + W(A^2)-U(x)A^2
\label{eq:Hamiltonian}
\end{equation}
with $W(A^{2}=n)=\frac{v_{\infty}^{2}}{2}\left( n+\frac{1}{n}\right)+ng(1)-G(n)$, and the antiderivative $G(n)=\int \mathrm{d}n \, g(n)$. Equations ~\eqref{eq:HydroEqs} are then derived from the canonical Hamilton equations $\dot p=-\partial_A \mathcal H$ and $\dot A=\partial_p \mathcal H$ where the dot stands for the total derivative with respect to the effective time $x$. In particular, in the absence of the external potential $U(x)$, this Hamiltonian is time-independent and the energy $E_{\mathrm{cl}}$ of the classical particle is conserved. The free solutions of the NLS equation can then be readily obtained from the possible trajectories of the classical particle in the potential $W(A^2=n)$. The typical shape of this potential for $g(n)=n$ is depicted in fig.~\ref{fig:Fict_Pot}. For example, the equilibrium point referred to as $n_{\mathrm{\infty}}$ in fig.~\ref{fig:Fict_Pot} corresponds to a constant-density supersonic solution ($v_\infty>c_\infty$) while small oscillations around this classical fixed point correspond to the superposition of an incoming plane wave and a small-amplitude reflected wave describing weak backscattering. Note that, in general, the nonlinearity of the NLS equation forbids such a separation of the upstream solution into a sum of independent incoming and reflected waves. However, if the backscattering is weak, the interaction between the incoming wave and the reflected wave is negligible. In general, this separation is not possible and the free solutions are described by cnoidal waves \cite{PavloffLeboeuf2001}. In particular, solutions with $E_\mathrm{cl}$ slightly below $W_\mathrm{max}$ correspond to one or several gray solitons. In the presence of the scattering potential $U(x)$ the energy of the classical particle is no longer conserved and its dynamics may be nontrivial. The boundary condition in the downstream region (where $U(x)=0$) imposes that the classical particles starts with $A(-\infty)=1$ ($n_\mathrm{\infty}$ in fig. \ref{fig:Fict_Pot}) and the forward integration has to satisfy that the final energy and the position $A$ of the classical particle remain in the well of $W(n)$. This is the strategy we use to obtain the equation of the supersonic separatrix.  

In the following we provide explicit analytical results for the supersonic separatrix in the limiting cases of weak, narrow and wide obstacles. We then focus on the case of the attractive obstacle of arbitrary width. Using numerical solutions, we identify resonant transport and solutions with perfect transmission similar to the ones of the superfluid regime. Our study reveals the existence of resonances for very specific sets of injection velocity and obstacle parameters as can be seen in fig.~\ref{fig:phase}, which we characterise in the context of our simplified model, providing a better comprehension of the phenomenon.

\section{Weak obstacle}
Before discussing nontrivial situations, it is instructive to revisit the Landau's criterion, which in our case reduces to the well known result $v_c = c_\infty$, in terms of the classical mechanics analogy. The fictitious potential $W(n)$ must be typically of the shape shown in fig.~\ref{fig:Fict_Pot}, with $\lim\limits_{n \to 0}W(n)=+\infty$ and $\lim\limits_{n \to +\infty}W(n)=-\infty$. $W(n)$ has a local minimum $W_{\mathrm{min}}$ obtained at $n_{\mathrm{\infty}}=1$, and a local maximum $W_{\mathrm{max}}$ for $n_{\mathrm{max}}>1$. In the absence of any scattering potential, solutions with $n=n_\infty$ and $n=n_\mathrm{max}$ correspond respectively to the supersonic and the subsonic superfluid solution. As $v_\infty$ is tuned and approaches $c_\infty$ these two solutions merge and the potential has a saddle point as shown in the inset of fig.~\ref{fig:Fict_Pot}. The two critical velocities $v_c$ and $v_s$ are therefore identical and equal to the speed of sound as shown in fig.~\ref{fig:phase} for $U_0\to 0$. The presence of a weak obstacle will not modify the structure of this saddle-node bifurcation but will only be the trigger of the instability.

\section{Narrow obstacle}
\label{sec:NarrowObstacle}
When the typical range of the obstacle potential is much smaller than the healing length, i.e. when $\sigma\ll 1$, it is possible to approximate $U(x)$ by $U(x)=U_{0}F(\sigma)\delta(x)$, where $F(\sigma)$ is the integral of $f(|x|/\sigma)$ over the whole real axis, and is simply given by $F(\sigma)=\sigma$ in the case of a square obstacle, which could be a well or a barrier depending on the sign of $U_0$. One can then obtain an analytical expression for the supersonic separatrix by searching for the solutions of the Hamilton equations with energy $E_{\mathrm{cl,\delta}}=\varepsilon(v_{\infty})=2U_{0}^{2}F^{2}(\sigma)+v_{\infty}^{2}+g(1)-G(1)$ associated to a $\delta$--shaped obstacle.

From a classical point of view, the fictitious particle starts at $x=-\infty$ with density $n_{\mathrm{\infty}}=1$. It experiences a kick of energy when meeting the obstacle, going from $W_{\mathrm{min}}$ to $\varepsilon$, and will oscillate between the two solutions of $W(n)=\varepsilon$ after this encounter. If $W_{\mathrm{min}}<\varepsilon<W_{\mathrm{max}}$, the particle is trapped and the density oscillates between the two solutions of $W(n)=\varepsilon$: This is the supersonic stationary regime. This type of solution is depicted in subfigure (c) of fig.~\ref{fig:phase}. However if $\varepsilon<W_{\mathrm{min}}$ or $\varepsilon>W_{\mathrm{max}}$, the dynamics is no longer stationary and excitations are continuously generated as depicted in subfigure (b) of fig.~\ref{fig:phase}.

The boundary between the nonstationary and the stationary regimes is by definition the supersonic separatrix, and corresponds to the last stationary solution. It is given by $\varepsilon(v_s)=W\left(n_{\mathrm{max}}(v_s)\right)$ with $n_{\mathrm{max}}$ such that $W'(n_{\mathrm{max}})=0$. This yields
\begin{equation}
\begin{aligned}
\frac{1}{\sqrt{2}}\bigg[\frac{v_s^{2}}{2}\bigg(\sqrt{n_{\mathrm{max}}}&-\frac{1}{\sqrt{n_{\mathrm{max}}}}\bigg)^{2}+g(1)(n_{\mathrm{max}}-1)\\
&{+}\,G(1)-G(n_{\mathrm{max}})\bigg]^{\frac{1}{2}}=|U_{0}F(\sigma)|.
\label{eq:separatrix_delta}
\end{aligned}
\end{equation}
An explicit solution of this equation can be derived for a cubic nonlinearity of the form $g(n)=n$ \cite{PavloffLeboeuf2001} while for a saturable nonlinearity of the form $g(n)=(1+n_{s})^2 n / \left[n_{s}(n+n_{s})\right]$, characteristic of superfluids of light in a saturable media \cite{Eloy2021}, it has to be solved numerically (see Supp. Mat.).

\begin{figure}[!t]
\centering
\includegraphics[width=.95\linewidth]{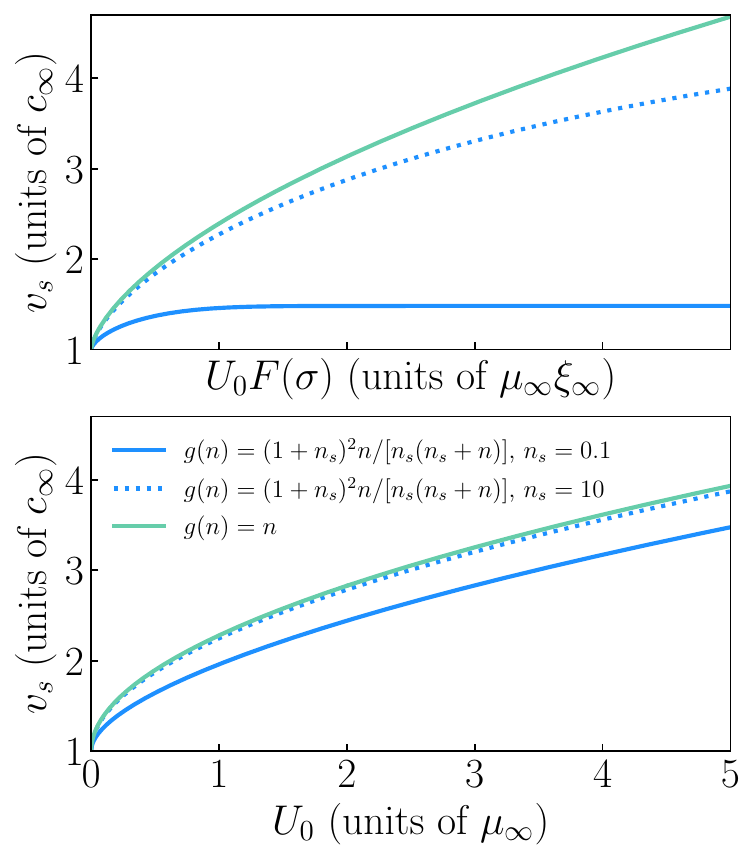}
\caption{\label{fig:delta}The top (bottom) figure represents the supersonic separatrix for a repulsive narrow (wide) obstacle of amplitude $U_{0}F(\sigma)$ ($U_{0}$). The two types of nonlinearity $g(n)$ considered here are indicated in the legend which applies to both figures.}
\end{figure}

The upper panel of fig.~\ref{fig:delta} represents the supersonic separatrix with respect to the effective amplitude $U_0 F(\sigma)$ of the narrow obstacle. The green curve is obtained for a cubic nonlinearity, whereas the blue plain (dotted) lines are for a saturable nonlinearity, with saturation intensity $n_{s}=0.1,\;10$. Although the cubic nonlinearity is a limiting case of the saturable nonlinearity when $n_{s}\gg n$, large deviations are observable even for $n_s=10$ which are of great importance for experiments with fluids of light. Moreover, it is important to emphasise that in the above-mentioned saturable systems, the fictitious potential $W(n)$ may be such that it has no local maximum depending on the value of $n_{s}$. With a saturable nonlinearity, when $v_\infty  > \sqrt{2+2 n_{s}}$, the potential $W$ has only one minimum and diverges towards $+\infty$ for both $n\rightarrow 0$ and $n\rightarrow +\infty$ (see Supp.~Mat.). The fictitious particle is then always trapped in this potential and all solutions for $v_\infty  > \sqrt{2+2 n_{s}}$ are stationary. This explains the plateau at large $U_0F(\sigma)$ in the top of fig.~\ref{fig:delta} for the $n_s=0.1$ case with a value 
$v_s = \sqrt{2.2} \simeq 1.48$. 
It is also interesting to note that eq.~\eqref{eq:separatrix_delta} predicts a symmetric supersonic separatrix as a function of $U_0$. This symmetry between repulsive and attractive obstacle -- not present in the case of the superfluid separatrix -- is also a peculiarity of the $\delta$--peak model, and will be broken as $\sigma$ increases, or in other words when the velocity of the flow is large enough so that the associated de Broglie wavelength is small enough to resolve the details of the potential. This can be seen, for example, in fig.~\ref{fig:phase} where the symmetry is clearly broken and resonances appear in the attractive case for a square potential of width $\sigma=1$ and a cubic nonlinearity.

\section{Wide repulsive obstacle}
\label{sec:WideRepulsiveObstacle}

We now consider the obstacle dependence of the separatrix in the case of a wide obstacle $\sigma\gg 1$. In that case, the fictitious particle follows adiabatically the slow variations of the effective potential $W(n)-U(x) n$ and remains at its minimum as far as it exists. The situation is then similar to the one of a weak obstacle. The fluid flow is locally uniform but dressed by the obstacle according to eq.~\eqref{eq:HydroEqs} with $A''=0$. It is now a matter of applying the criterion for the saddle-node bifurcation locally, namely looking for points where the local velocity is equal to the local speed of sound $c(x)=n(x)g'(n(x))$. This will first appear at the maximum of the scattering potential at $x=0$. Using current conservation we obtain the implicit formula  

\begin{equation}
\label{eq:N0cWide0}
g'(n_{0,\mathrm{c}}^{\vphantom{3}})n_{0,\mathrm{c}}^{3}=v_s^{2},
\end{equation}
where $n_{0,c}$ is the density at $x=0$ at the transition point and can be computed from the saddle point equation $W'(n_{0,c})-U_0=0$. Equation~\eqref{eq:N0cWide0} is equivalent to $W''(n_{0,c})=0$ which is the second saddle point equation. It is important to note that this equation is the same for the supersonic and the subsonic separatrix but has two solutions respectively larger and smaller than $c_\infty$. As for the narrow barrier, this supersonic separatrix is represented in the bottom part of fig.~\ref{fig:delta} for an obstacle of typically large $\sigma$, and for the same kinds of nonlinearities. As a matter of fact, the effect of saturation of the nonlinearity is less pronounced for large obstacles than for narrow ones. Note that we do not display the attractive part since eq.~\eqref{eq:N0cWide0} predicts that the critical velocity is always the sound velocity at this level of approximation.

\section{Attractive obstacle of arbitrary width}
\label{sec:ArbitraryWidth}

\begin{figure}[!t]
\centering
\includegraphics[width=.96\linewidth]{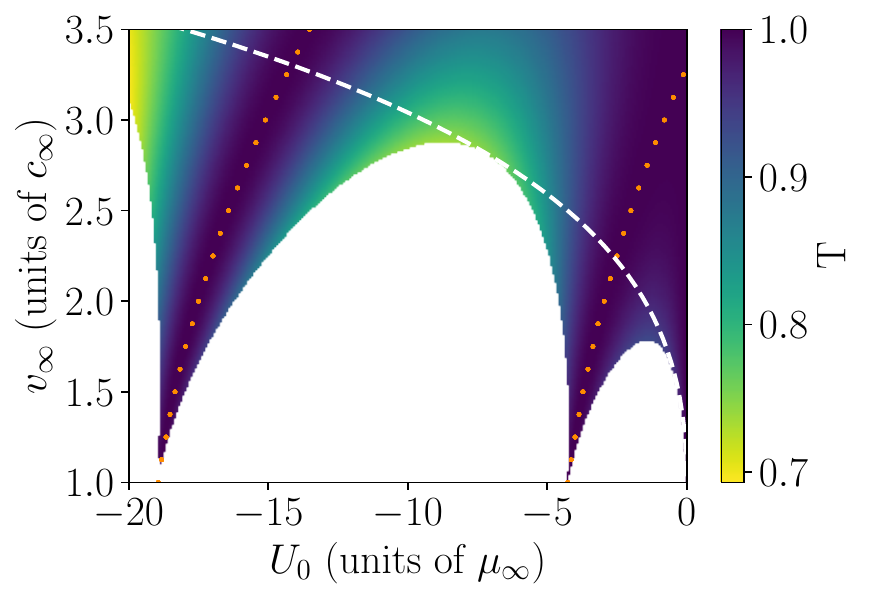}
\includegraphics[width=.96\linewidth]{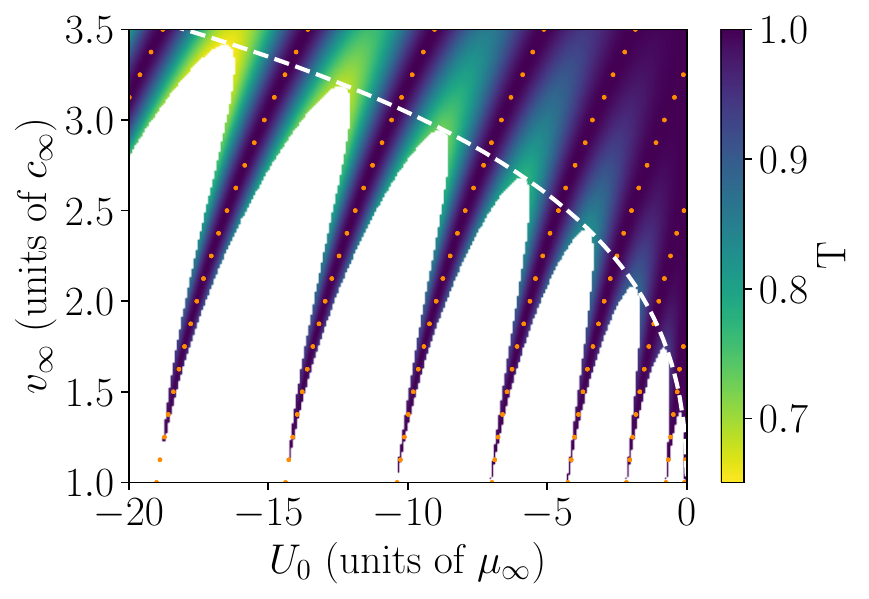}
\caption{Phase diagram $(U_{0},v_{\infty})$ (in the natural units of the superfluid) of a quantum fluid flowing across attractive square well potentials of respective width of $\sigma=1$ (top) and $\sigma=4$ (bottom), and for a cubic nonlinearity of the form $g(n)=n$. The transmission across the barrier is associated with the colour bar and is maximum along the orange dotted curves which determine the position of the resonances, whereas the white dashed line represents the envelope of said resonances.}
\label{fig:Transmission}
\end{figure}

In the general case, the precise shape of the obstacle has an important influence as we will discuss in this section. In general, eq.~\eqref{eq:HydroEqs} has to be solved numerically to obtain the equation of the separatrix, except for specific models such as piece-wise constant obstacles \cite{PavloffLeboeuf2001}. However, as far as localised obstacles of the form discussed in this work are considered, the generic picture displayed in fig.~\ref{fig:phase} is valid. In particular, nonlinear resonances may exists and lead to a nontrivial structure of the stability diagram. In ref. \cite{ParisMandoki2016} such resonances were considered in the case of a repulsive square well obstacle due to the Ramsauer-Townsend effect in arbitrary dimension. These solutions were put forward to be of great interest since they share an important property with superfluid solutions, and they do not experience friction with the obstacle although they are supersonic (see Supp. Mat. for a detailed analysis). However, they exist on specific curves (represented by the orange dotted curves in fig. \ref{fig:Transmission}) in the stability diagram and do not form a continuous family of solutions like the subsonic superfluid solutions. We then cannot find a real superfluid regime above the supersonic separatrix as these lines form a null measure set. In the following, we discuss in details the case of an attractive potential and give explicit results for a square well potential. Results with a Gaussian potential are available in the Supp. Mat. In particular we demonstrate that the lobe structure in the stability diagram of fig. \ref{fig:phase} is indeed related to these resonances which continuously connect the superfluid solutions to superfluid-like solution above the critical velocity along one-dimensional lines in parameter space $(U_0, v_\infty)$.

From now on, we focus on the attractive case and exemplify our findings with a square well potential of amplitude $U_0$ and width $\sigma$, and complement the stability diagram with the knowledge of the transmission coefficient in the $(U_0,v_\infty)$ plane. While in the linear case [i.e. $g(n)=0$ in eq.~\eqref{eq:GNLSE}] the reflection and transmission coefficients, as well as the position of the resonances, are well known \cite{Griffiths2018}, they cannot be defined easily in the nonlinear case as previously discussed. Nevertheless it is possible to give a proper definition of scattering amplitudes using the theory of adiabatic invariants \cite{LandauVol1,Paul2007} or a simpler but perturbative one in the weak backscattering limit \cite{Paul2007,Paul2009}. As we are mostly interested in the position of the resonances, we will employ the latter definition which has been shown to produce results in good agreement with the exact 
adiabatic invariants approach \cite{Paul2007} even far away from resonances. The transmission coefficient reads
\begin{equation}
T=\left(1+\frac{\Delta E}{2(v_{\infty}^{2}-1)}\right)^{-1}.
\end{equation}
$\Delta E$ is the energy difference of the fictitious particle between its final 
($x=+\infty$) and initial ($x=-\infty$) states: $\Delta E = \mathcal H\left[A(x=+\infty),p(x=+\infty)\right]-\mathcal H\left[A(x=-\infty),p(x=-\infty)\right]$.
For $x=-\infty$, the fictitious particle
is at equilibrium with $A=1$ and $p=0$.

\begin{figure}[!t]
\centering
\includegraphics[width=.95\linewidth]{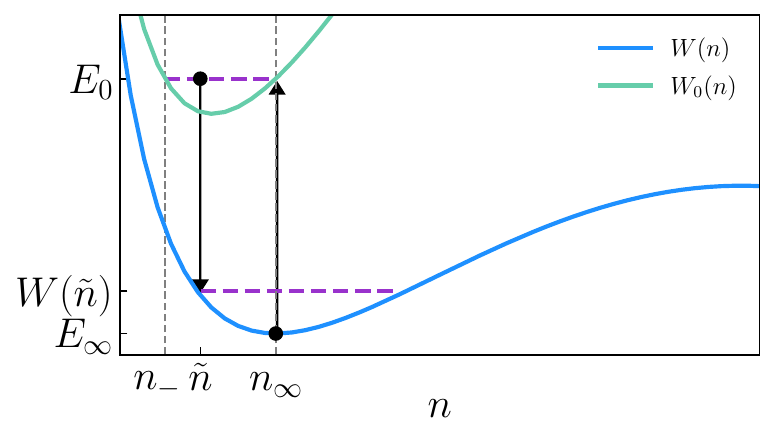}
\caption{Classical potentials seen by the fictitious particle in the case of an attractive square well obstacle. The lower curve is the potential  $W(n)$ for $x<-{\sigma}/{2}$ and $x>{\sigma}/{2}$ while $W_0(n)$ is the one for $x\in[-{\sigma}/{2},{\sigma}/{2}]$.}
\label{fig:cas1}
\end{figure}

Numerical results are summarised in fig.~\ref{fig:Transmission}. The colour scale shows the transmission coefficient of the fluid across the obstacle as a function of the injection velocity of the fluid and of the amplitude of the square well obstacle, for a given value of $\sigma$. The coloured zone is separated from white zones of undefined transmission (corresponding to the nonstationary regime) by the supersonic separatrix, clearly exhibiting resonances. In particular, the perfect transmission lines are shown to follow exactly the nontrivial structure of the stability diagram and are drawn as orange dotted curves, while the white dashed curve represents the envelope of the resonances. Both curves can be calculated analytically for a square well potential as suggested in \cite{PavloffLeboeuf2001} and explained below. In the following we provide explicit results for $g(n)=n$. Again, thinking in terms of a fictitious particle moving in a classical potential provides a simple picture of the underlying physics, and the mechanism behind the existence of resonances is illustrated in fig.~\ref{fig:cas1}.

We start by discussing the stability diagram. Before the excitation caused by the rectangular obstacle, the fictitious particle is at rest from $x=-\infty$ to $x=-\sigma/2$ with density $n_{\infty}$ and energy $E_{\infty}=W(n_{\infty})$ in the potential $W$. As it reaches the obstacle, it undergoes a kick of energy $\Delta E = E_{0}-E_{\infty}$, going from $(n_{\infty},E_{\infty})$ to $(n_{\infty},E_{0})$ in the new potential $W_{0}=W+U_{0}n$. The particle then oscillates in $W_{0}$ between $n_{\infty}$ and $n_{-}$ as it progresses in the obstacle, and returns to the potential $W$ with density $\Tilde{n}$ for $x=+\sigma/2$. Several cases leading to different dynamics for the fluid are then possible depending on the values of $U_{0}$, $v_{\infty}$ and $\sigma$. In that context, we define $\Tilde{L}$ the distance between $n_{\infty}$ and $\Tilde{n}$ performed in $W_{0}$, and $L_{0}$ the distance of the round-trip between $n_{\infty}$ and $n_{-}$, i.e. the period of the oscillations of the fictitious particle in $W_{0}$

\begin{subequations}\label{eqperiod}
\begin{equation}
     \Tilde{L} = \frac{1}{\sqrt{2}}\int^{\sqrt{n_{\infty}}}_{\sqrt{\Tilde{n}}}\frac{\mathrm{d}A}{\sqrt{E_{0}-W_{0}(A)}} ,  
\end{equation}
\begin{equation}
        L_{0} = \sqrt{2}\int^{\sqrt{n_{\infty}}}_{\sqrt{n_{-}}}\frac{\mathrm{d}A}{\sqrt{E_{0}-W_{0}(A)}}.
\end{equation}
\end{subequations}

For stationary solutions to exist it is necessary that the energy of the fictitious particle when it exits the obstacle is lower than the maximum of $W(n)$ (the configuration of fig.~\ref{fig:cas1}). That way, the particle is always confined. The envelope of the resonances (white dashed line in fig.~\ref{fig:Transmission}), above which stationary solutions exist no matter the value of $\sigma$, is obtained when the energy of the fictitious particle at the end of the obstacle corresponds exactly to the maximum of the fictitious potential. An analytical expression can be obtained for that envelope, separating the case where solutions always exists from the one where the existence of said solutions depends on the value of $\sigma$, and can be found in eqns.~(34) and (35) of \cite{PavloffLeboeuf2001}. 
It is interesting to note that, for a square well obstacle, $U_{0}$ and $\sigma$ are uncorrelated quantities, and the amplitude of the resonances will not depend on $\sigma$ as shown by the white dashed line in fig.~\ref{fig:Transmission}. For such an obstacle, the resonances will never disappear and their envelope will always be the same for any value of $\sigma$. Interestingly, numerical simulations showed that this is not the case for a Gaussian potential: The envelope of the resonances does depend on $\sigma$, and decreases as the width increases. These results are presented in the Supp. Mat. One can also see in fig.~\ref{fig:Transmission} that the resonances multiply as $\sigma$ increases. At some point, for an arbitrarily large value of $\sigma$, the resonances are so thin and numerous that they are not distinguishable from one another anymore, to the extent that the supersonic separatrix would be given by $v_s=1$ in the limit $\sigma\gg1$.
Since the subsonic separatrix is also given by $v_c=1$, the gap opened by the nonstationary regime slowly closes as the width of the obstacle increases. 

Concerning the position of the resonances, the connection of the subsonic superfluid solution to curves of perfect transmission in the supersonic regime is performed at $v_{\infty}=c_{\infty}$. From our classical analysis, the two extrema of the potential $W(n)$ merge in a unique saddle point located at $n=n_{\infty}$. Oscillations are not possible anymore, and the only way for a stationary state to exist is when the excited fictitious particle exits $W_{0}$ with the same density it had when entering it, meaning $\Tilde{L}=0$. More generally, when the fictitious particle performs an arbitrary number of round-trips in the excited potential $W_{0}$, so that its energy after exiting the obstacle is exactly the one it had before the excitation, a resonance forms between the width of the obstacle and the wavelength of the cnoidal wave of the oscillating particle, causing a perfect transmission, and linking the superfluid regime to the stationary nonsuperfluid one. The equation of theses lines of perfect transmission (see the orange dotted curves in fig.~\ref{fig:Transmission}) is then given by $\alpha L_{0}=\sigma$, $\alpha$ being an integer, with $L_0$ given by eq.~\eqref{eqperiod}. Along these lines, the superfluid/stationary nonsuperfluid transition is continuous and the system is always stationary. Note that we have numerically checked the stability of this solution by doing time-dependent simulation of the NLS eqs.~\eqref{eq:GNLSE}. This was done in the spirit of ref. \cite{ParisMandoki2016} where the obstacle was ramped adiabatically in strength and velocity.

\section{Conclusion}
In this paper we have studied the supersonic separatrix between the nonstationary and the normal stationary regime of a generic quantum fluid flowing past a localised obstacle of arbitrary amplitude and width, in the 1D mean-field regime. We have computed this critical velocity by deriving nonperturbative exact expressions in the limits of narrow- and wide-obstacle, and studying more thoroughly the peculiar case of the arbitrary attractive obstacle, which exhibits a nontrivial behaviour. For most parameters, a standing wave forms ahead of the obstacle, with a constant friction force and a partially transmitted wave. However we have shown that, along certain curves, due to resonant transport, supersonic solutions may share fundamental properties with their subsonic superfluid counterpart: They are symmetric solutions and, therefore, the quantum fluid is totally transmitted through the obstacle, without any friction.

All together, these results are important for the experimental studies of transport properties of quantum fluids described by a generalised NLS equation. They provide a clear map in parameter space of the different possible stationary regimes of flow for a quantum fluid, in order to guide experimental studies in the desired regime of nonlinear transport.

Finally, an extension of this work to higher dimensions would be desirable, notably in 2D, as experimental data are available for a saturable nonlinearity \cite{Eloy2021}.

\acknowledgments
We acknowledge P. Vignolo, M. Bellec and C. Michel for inspiring discussions. This work has benefited from the financial support of Agence Nationale de la Recherche under Grants Nos.~ANR-21-CE30-0008 STLight (Superfluid and Turbulent Light in Complex Media) and ANR-21-CE47-0009 Quantum-SOPHA (Quantum Simulators for One-Dimensional Systems with Photons and Atoms).

\end{document}


\maketitle

\section{Nonlinear Schr\"odinger equation dictionary} \label{sec:NLSdict}

The mean-field dynamics of various 1D systems can be described by generalised NLS equations. Two instances of such systems, exemplified in the main paper, will be given in this section.

This equation is mainly known to describe the dynamics of the one-dimensional reduction $\psi(x,t)$ of the condensate wave function of a dilute ultracold atomic Bose gas.
In that case, the system consists of weakly repulsive identical atoms in a highly asymmetric harmonic trap, which makes the evolution of the condensate to be quasi one-dimensional along a given axis of the trap \cite{Pitaevskii2016}.
In the present paper, we mostly work with a toy model consisting of a square  potential for simplicity but, from an experimental point of view, the obstacle potential $U(x)$ can be realised by crossing the atomic cloud with a detuned laser beam, larger than the transverse size of the condensate \cite{Engels2007}.
The flow of the Bose fluid in a given direction can then be simulated by displacing the laser beam creating the obstacle in the opposite direction \cite{Engels2007}, which is equivalent to looking at the system in a reference frame where the fluid is at rest.
In that context, the self-interaction potential $g(n)$ of the condensate is proportional to $n$ and given by $g(n)=2\hbar \omega_\perp n a_s$ where $a_s$ is the s-wave scattering length of the two-body interaction potential, $\omega_\perp$ is the transverse frequency of the harmonic trap, and the system is
dilute $n a_s \ll 1$. This configuration of a dilute gas is the one exemplified in the main text but in its dimensionless form $g(n)=n$, obtained after a relevant rescaling of the main quantities.

The NLS equation is also used in the optics domain, for example to describe the propagation of a scalar laser field in a local nonlinear medium. Such interactions between light and matter can be encountered in various domains pertaining to nonlinear optics or atomic physics \cite{Agrawal,Boyd,Vocke2016,Michel2018,Eloy2021,Landau1941a,Santic2018,Larre2015,Leboeuf2010}.
A particular realisation, relevant to the study of the transport of a fluid of light around an obstacle, is realised by the paraxial propagation of a monochromatic optical ﬁeld in a nonlinear medium \cite{Michel2018,Eloy2021,Carusotto2013}.  Such systems can be mapped onto a one-dimensional Gross-Pitaevskii-type evolution
of a quantum ﬂuid of interacting photons in the plane transverse to the propagation \cite{Pitaevskii2016}, the propagation coordinate playing the role of time. The transverse direction represents the space in which the fluid of light evolves, which is generally two-dimensional. An obstacle $U(x)$ can be introduced through a spatial modulation in the linear refractive index of the medium \cite{Michel2018,Eloy2021}. The effective mass $m$ is related to the propagation constant of the fluid-of-light beam propagating in the medium, the density of the fluid is given by the light intensity, and its velocity corresponds to the gradient of the phase of the optical field.
The photon–photon interactions, mediated by the nonlinear response of the material in which the fluid of light is propagating, lead to different kinds of nonlinearities, depending on the medium that is considered.
For example, a defocusing Kerr medium gives a nonlinearity $g(n)$ that increases linearly with the light intensity $n$ \cite{Agrawal,Boyd}. In a defocusing saturable medium, like the nonlinear photorefractive crystal used in the experiments of refs.~\cite{Michel2018, Eloy2021}, the nonlinearity takes the saturable form $g(n)=\pi N^{3}r_{33}E_{0}n/[\lambda_{0}(n_{\mathrm{s}}+n)]$, where $N$ and $r_{33}$ are respectively the mean refractive index and the electro-optic coefficient of the crystal along the extraordinary axis. $E_{0}$ is the amplitude of an electric field applied to the crystal, $\lambda_{0}$ is the wavelength of the laser carrier in free space, and $n_{\mathrm{s}}$ is a saturation intensity adjusted by illuminating the crystal with white light.
In the present paper, the dimensionless version of the saturable nonlinearity reads $g(n)=(1+n_s)^2 n/\left[n_s(n_s+n)\right]$, and is once again obtained through the proper rescaled units.

\section{Behaviour of the fictitious potential for a saturable nonlinearity}
\begin{figure}
\centering
\includegraphics[width=\linewidth]{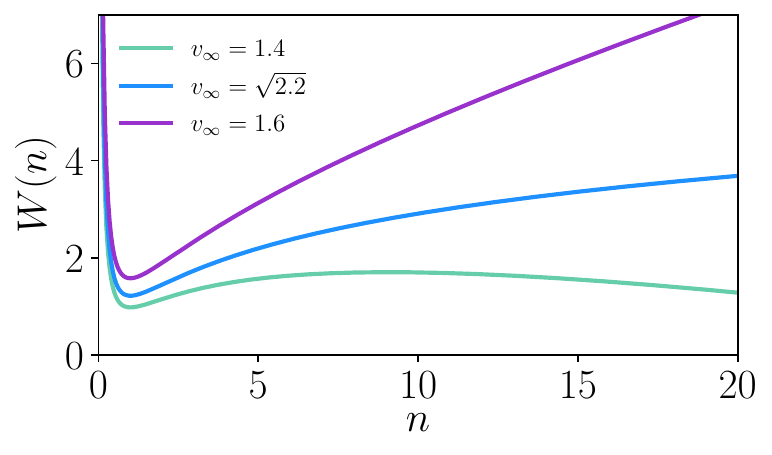}
\caption{\label{fig:sat}The fictitious potential $W(n)$ in the case of a saturable nonlinearity for $n_s=0.1$ and for different velocities $v_\infty$. The behaviour in the limit $n\rightarrow \infty$ changes for $v_\infty=\sqrt{2+2n_s} \simeq 1.48$ for $n_s=0.1$, diverging towards $-\infty$ for $v_\infty<\sqrt{2+2n_s}$ and towards $+\infty$ for $v_\infty\ge\sqrt{2+2n_s}$.
}
\end{figure}

In the case of a saturable nonlinearity, the
fictitious potential $W(n)$ shows a behaviour that can be qualitatively different from the behaviour observed for a cubic nonlinearity (illustrated by fig.~2 and fig.~5 of the main article). Figure \ref{fig:sat} shows that, at low velocity $v_\infty$, $W(n)$ shows a minimum and a maximum at finite values and diverges towards $-\infty$ for $n\rightarrow +\infty$. In that case, the behaviour is similar to the case of a cubic nonlinearity.

However, for $v_\infty \ge \sqrt{2+2n_s}$, $W(n)$  diverges towards $+\infty$ for $n\rightarrow +\infty$ and shows only one minimum. This divergence is logarithmic at the threshold and linear above it. The fictitious particle is then always trapped in a potential well, whatever its energy. This means that the fictitious particle is always stable or oscillating around this minimum and there are only stationary solutions for $v_\infty \ge\sqrt{2+2n_s}$. This translates into the saturation of the supersonic separatrix $v_s$ observed in fig.~3 of the main article at large $U_0F(\sigma)$.

\section{Explicit expressions for the critical velocity}
Based on previous studies \cite{PavloffLeboeuf2001,Hakim1997,Pavloff2002}, we first detail the calculations leading to the expression of the upper separatrix in the case of an obstacle potential of the form $U(x)=U_0F(\sigma)\delta(x)$, in the limit $\sigma\ll1$.
The equation to obtain the critical velocity, derived from eq. (2) of the main paper, reads
\begin{equation}
\frac{1}{2}\frac{\partial_{xx}\sqrt{n}}{\sqrt{n}}+\frac{v_{\infty}^{2}}{2}\bigg(1-\frac{1}{n^{2}}\bigg)+g(1)-g(n)=U_0 F(\sigma)\delta(x).
\end{equation}
The process of solving this equation for the density can be separated into two parts: First, for $x=0$, which yields a condition on the first derivative of the density. This condition will then act as a link between the two solutions for $x<0$ and $x>0$.
The radiation condition \cite{Lamb1997} imposes that $n(x)=n_\infty$ in the region upstream of the obstacle, which forces $\partial_x n(0^-)=0$. In the downstream region, an infinite cnoidal wave is generated, for which the expression of the density is given by the solution of
\begin{equation}
\label{Eq:picdelta}
\frac{1}{2}\frac{\partial_{xx}\sqrt{n}}{\sqrt{n}}+\frac{v_{\infty}^{2}}{2}\bigg(1-\frac{1}{n^{2}}\bigg)+g(1)-g(n)=0.
\end{equation}
A graphic representation of the upstream and downstream solutions for the density, obtained after a numerical integration of the Gross-Pitaevskii equation, can be found in fig. \ref{fig:density_res}.

It is possible to obtain exact analytical results for the upper separatrix. The procedure to do so is pretty straightforward: Obtain the expression of $n_{\mathrm{max}}$ through $W'(n_{\mathrm{max}})=0$, and then inject it in $\varepsilon(v_s)=W(n_{\mathrm{max}}(v_s))$, which corresponds to eqn. (4) from the main paper.
For a cubic nonlinearity, we find
\begin{equation}
    n_{\mathrm{max}}(v_s)=\frac{v_s^{2}+v_s\sqrt{v_s^{2}+8}}{4},
\end{equation}
which then leads to
\begin{equation}
    |U_{0}F(\sigma)| = \frac{\sqrt{v_s(v_s^{2}+8)^{3/2} +v_s^{4}-20v_s^{2}-8}}{4\sqrt{2}}.
    \label{eq:separatrix_delta}
\end{equation}
On the other hand, things become quite cumbersome when considering a saturable nonlinearity, and even if it is possible to obtain $n_{\mathrm{max}}$ analytically

\begin{equation}
    n_{\mathrm{max}}(v_{s}) = \frac{v_{s}^{2}(1+n_{s})+v_{s}\sqrt{8n_{s}(1+n_{s})+v_{s}^{2}(1-n_{s})^{2
    }}}{2(2+2n_{s}-v_{s}^{2})},
\end{equation}
the equation leading to the critical velocity has to be solved numerically.

In the wide-obstacle limit $\sigma\gg1$, the gradients of $U(x)$ are small enough that the fluid behaves as if it were uniform. Based on a rigorous multiple-scale treatment of the obstacle potential when $\sigma\gg1$, we consider an obstacle of the form $U(x)=U_0 \left[ 1+f''(0)x^2/(2\sigma^2)\right]$, where the terms in the square brackets correspond to the series expansion of $f\left( |x/\sigma| \right)$ to second order in $1/\sigma\ll1$.
The method to obtain the equation to the supersonic separatrix is the following: The last existing solution to
\begin{equation}
\frac{v_{\infty}^{2}}{2}\bigg(1-\frac{1}{n^{2}}\bigg)+g(1)-g(n)=U_0
\end{equation}
(eq. (2) from the main paper with a flat obstacle of amplitude $U_0$, i.e. the zeroth-order in the $1/\sigma$ development of the obstacle potential) is obtained for $v_\infty=v_s$ and $n=n_{0,\mathrm{c}}$, defined through the condition $g'(n_{0,\mathrm{c}})n_{0,\mathrm{c}}^3=v_s^2$.
Exact analytical results were obtained for the critical velocity with respect to the amplitude of the obstacle (as its width is fixed and supposed large), and for a cubic nonlinearity

\begin{equation}
\begin{aligned}
v_{s,0}=\Bigg[2(U_0-1)&+3\left(U_0-1 + \sqrt{U_0 (U_0 -2)}\right)^{-\frac{1}{3}}\\
&{+}\,3\left(U_0-1 + \sqrt{U_0 (U_0 -2)}\right)^{\frac{1}{3}} \Bigg]^{\frac{1}{2}}.
\label{eq:separatrix_large}
\end{aligned}
\end{equation}
Similar results were obtained numerically for a saturable nonlinearity.
Contrary to the critical velocity for superfluidity (i.e. the lower separatrix) for which superfluidity is broken for a given $U_{0,\mathrm{max}}$ which depends on the chosen nonlinearity \cite{Huynh2022}, eq. \eqref{eq:separatrix_large} is valid for any amplitude of the obstacle. However, the bigger $U_0$, the harder it will be to reach a stationary regime, which is pretty intuitive as an important obstacle will induce more nonlinear emissions inside the fluid than a relatively small one, and the regime of quantum turbulence will be harder to depart from.

\begin{figure}[!t]
\centering
\includegraphics[width=\linewidth]{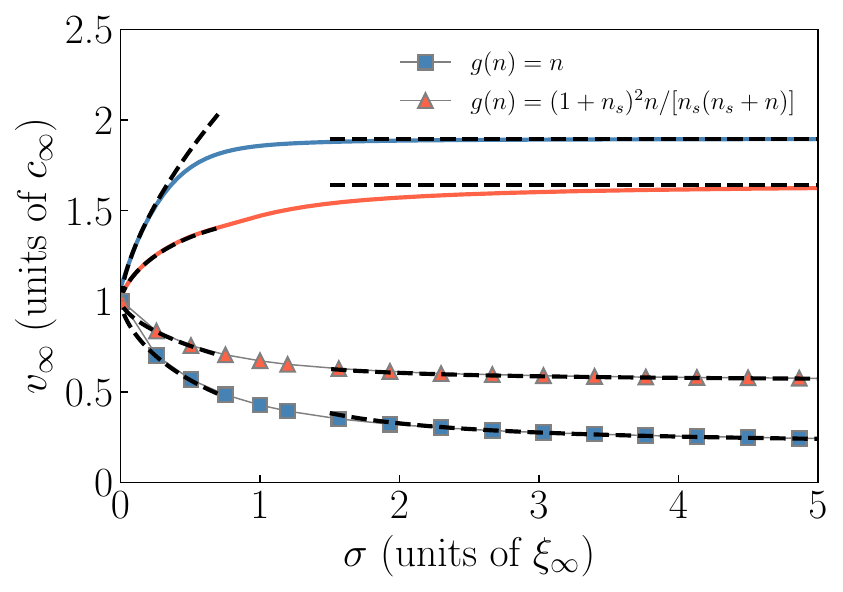}
\caption{\label{fig:phasegauss}Critical velocities $v_c$ or $v_s$ in the subsonic and supersonic regimes as a function of the typical range $\sigma$ of the obstacle potential $U(x)=U_0 \exp{(-x^2/\sigma^2)}$ supposed to be repulsive. The velocities are plotted at a fixed $U_0=0.5$ for two differents $g(n)$'s. The asymptotic results for the $\delta$--limit and the wide obstacle limit are represented in black dashed lines and are in relatively good agreement with numerics. Plain colored lines correspond to the supersonic separatrix $v_s$ and lines with symbols to the superfluid critical velocity $v_c$.}
\end{figure}

The approach used in \cite{Hakim1997,Huynh2022} to perform an analytical treatment of the higher order in $1/\sigma\ll1$ is no longer valid for the supersonic separatrix: It assumes that $n(x)=n_{0,c}+\delta n(x)$ with $\delta n(x)/n_{0,c}\ll1$, meaning small density fluctuations around the obstacle, which is clearly not the case as $v_\infty$ has no upper bound and can lead to density fluctuations of any amplitude.
However, a numerical simulation showed that if it exists, this correction to the zeroth-order is extremely small compared to $v_{s,0}$. Given these results, we consider the hydraulic approximation sufficient to describe the supersonic separatrix in the wide-obstacle limit.

\section{Results for a Gaussian obstacle}
After obtaining results in the two limits of the wide and narrow obstacle, it is natural to move to the generic situation of a localised obstacle  of arbitrary range, for which the supersonic separatrix is not trivial.
We performed a numerical simulation yielding the results illustrated in fig. \ref{fig:phasegauss} for a repulsive Gaussian obstacle of the form $U(x)=U_0 \exp{(-x^2/\sigma^2)}$, with $U_0=0.5$ and for the two nonlinearities considered before: $g(n)=n$ and $g(n)=(1+n_s)^2n/[n_s(n_s+n)]$. The upper part of the figure ($v_\infty>1$) depicts the supersonic separatrix $v_s$ as a function of the width of the obstacle (the focus of the main article), whereas the lower part ($v_\infty<1$) encompasses the previous results obtained in \cite{Huynh2022} for the critical velocity for superfluidity $v_c$.

\begin{figure}[!t]
\centering
\includegraphics[width=\linewidth]{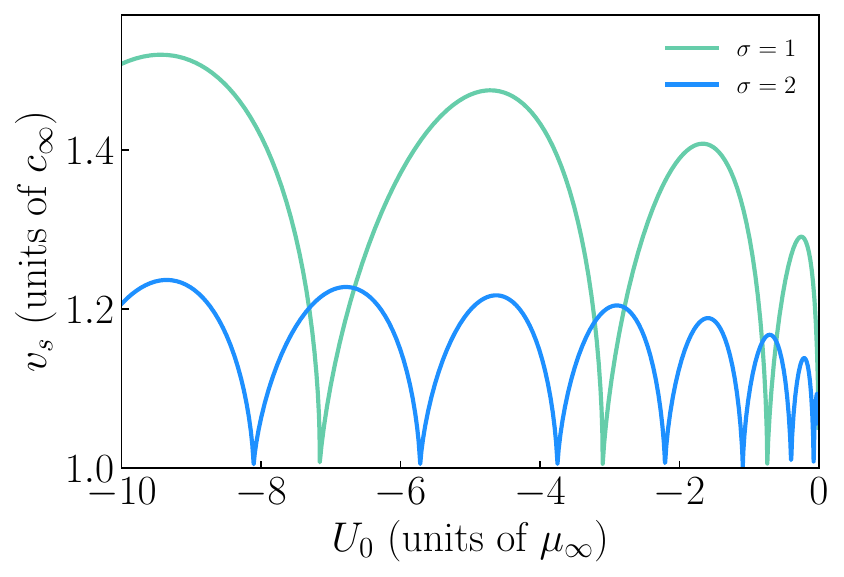}
\caption{\label{fig:resogauss}Supersonic separatrix for a Gaussian potential $U(x)=U_0 \exp{(-x^2/\sigma^2)}$ as a function of its negative amplitude. The results presented here are obtained through a numerical simulation for two different width $\sigma=1$ and $\sigma=2$, and clearly exhibit resonances for particular values of $U_0$.
}
\end{figure}

In addition, we raised the question of the attractive obstacle, for which nontrivial results were obtained after the numerical simulation we performed.
They indeed exhibit many resonances (except in the $\delta$--peak limit) that multiply as $\sigma$ increases, and that are delimited by an envelope which can be analytically determined in the context of our toy model (see the next section).
These results applied to a Gaussian obstacle exhibit another interesting feature. The envelope of said resonances gets a negative dependence on $\sigma$ as can be seen in fig. \ref{fig:resogauss} (which is not the case for a square well obstacle): The wider the obstacle, the more numerous the resonances and the lower their envelope.

\section{Characterisation of the resonances}

When looking at the phase diagram $(U_0,v_\infty)$, one can see that three different regimes coexist for an attractive obstacle: A nonstationary regime under the lobes, a regime that is always stationary located above the envelope of the resonances, and a regime that can be stationary depending on the value of $\sigma$, which is between the envelope and the lobes.
Going back to a square obstacle and following the calculations of \cite{PavloffLeboeuf2001}, we obtained an analytical expression for the envelope of the resonances, which can also be found in eqns. (34) and (35) of \cite{PavloffLeboeuf2001}
\begin{equation}
1-\frac{F(v_{\infty})}{2U_{0}}=G(v_{\infty},U_{0}),
\end{equation}
\begin{equation}
\begin{aligned}
F(v_{\infty})=&\left[\frac{v_{\infty}^{2}}{4}\left(1+\sqrt{1+\frac{8}{v_{\infty}^{2}}}\right)-1  \right]\\
&\times\left[\frac{5v_{\infty}^{2}}{4} +1-\frac{3v_{\infty}^{2}}{4}\sqrt{1+\frac{8}{v_{\infty}^{2}}}, \right]
\end{aligned}
\end{equation}
\begin{equation}
G(v_{\infty},U_{0})=\frac{v_{\infty}^{2}+1}{2}+U_{0}-\left[\left(\frac{v_{\infty}^{2}+1}{2}+U_{0} \right)^{2}-v_{\infty}^{2}\right]^{\frac{1}{2}}.
\end{equation}

\begin{figure}[!t]
\centering
\includegraphics[width=\linewidth]{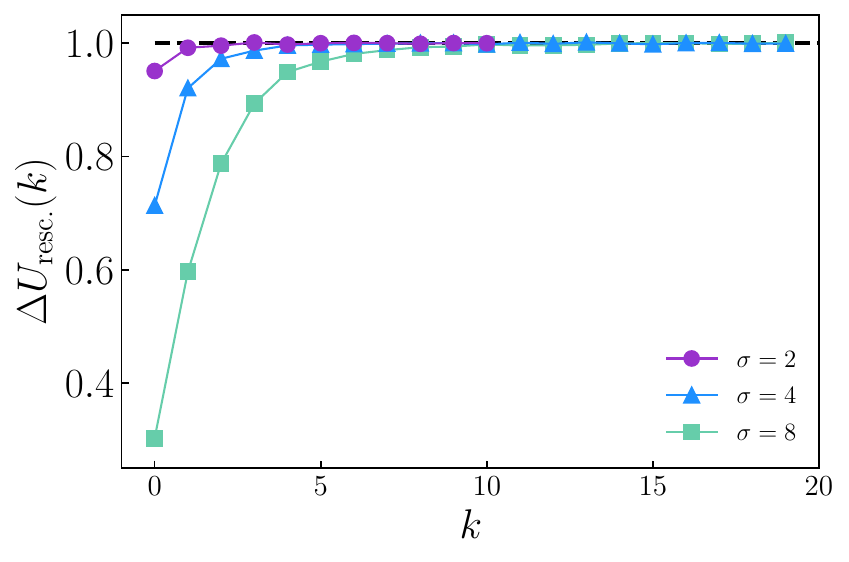}
\caption{Rescaled value of the separation between the resonances for a square well obstacle as a function of $k$ the index of the resonance. The coloured dots, triangles and square stand respectively for $\sigma=2$, $\sigma=4$ and $\sigma=8$ in the nonlinear regime $g(n)=n$ and are obtained after a numerical simulation, whereas the black dashed curve is the theoretical value for $g(n)=0$.}
\label{fig:espacement}
\end{figure}

Concerning the transmission coefficient, as $g(n)$ becomes non-null, the usual approach in terms of incident and reflected waves is not possible anymore because the problem is nonlinear, and the resonances, defined by the condition $T=1$, are slightly shifted as discussed in \cite{Paul2005,Rapedius2006}. We used a numerical simulation based on eq. (6) from the main paper to compute it, yielding fig.~4 of the main article with $T$ represented by the associated colour bar.

The aim is to characterise these resonances on the line $v_\infty=1$, as it creates unbroken lines of perfect transmission linking the superfluid regime to the stationary supersonic one.
To get rid of the energy offset, we look at the distance between the resonances and compare that to the linear case, for which the transmission coefficient is easily recovered \cite{Cohen} and in which case two consecutive resonances are separated by $(2k+1)\pi^2/(2\sigma^2)$.
We introduce the rescaled separation between each consecutive resonances
\begin{equation}
    \Delta U_{\mathrm{resc.}}(k)=\frac{2\sigma^2}{(2k+1)\pi^{2}}|U_{0,k+1}-U_{0,k}|,
    \label{eq:separation}
\end{equation}
and plot it in fig. \ref{fig:espacement}.
This function is constant (and equal to 1) in the linear case for $g(n)=0$ and is plotted in black dashed line, whereas the dots, triangles and squares are numerically obtained for different values of $\sigma$ in the nonlinear case.
As the left-hand side of eq. \eqref{eq:separation} gets rid of any $\sigma$--dependence, the different coloured curves should collapse on the black dashed one if the linear and the nonlinear case were following the same distribution.
One can see in fig. \ref{fig:espacement} that this is the case for large values of $k$.
However, there is a clear deviation from the linear case for small values of $k$, which is more important for wider obstacles.
The nonlinear effects then have a bigger impact on the separation between the resonances for small $k$ and for large $\sigma$.
It is also interesting to note that we have less data for $\sigma=2$ than for the other cases as this obstacle configuration leads to fewer resonances.

\begin{figure}[!t]
\centering
\includegraphics[width=\linewidth]{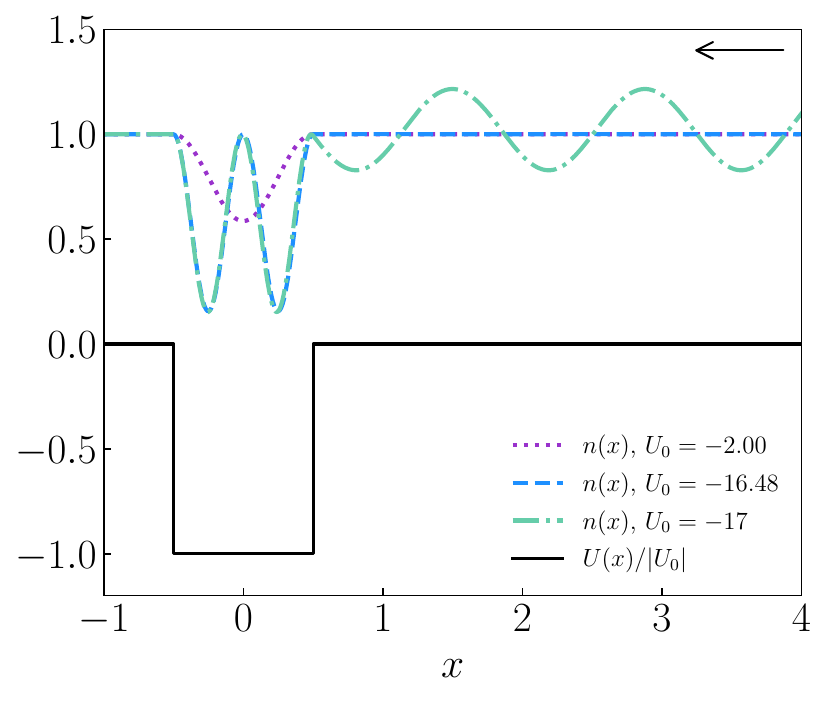}
\caption{Density profiles for several chosen parameters in the $(U_0,v_\infty)$ plane. The flow goes from right to left with constant velocity $v_\infty=2.5$, and encounters a square well obstacle of width $\sigma=1$.}
\label{fig:density_res}
\end{figure}

Concerning the superfluid-like solutions in the supersonic regime, we plot in fig. \ref{fig:density_res} the density profile for several points in the $(U_0,v_\infty)$ plane.
The dotted and dashed curves are respectively located close to the first and second line of perfect transmission ($k=1$ and $k=2$). The density is qualitatively the same as in the superfluid regime: A localised dip where the obstacle is, and a flat profile otherwise. The only difference is that as $k$ increases, so does the number of oscillations in the density profile. When the chosen parameters do not coincide with a resonance, the density profile is similar to the one in the supersonic regime when $U_0>0$, as represented with the dashed-dotted line.
This case can be better understood with an analogy with the linear case: A wave coming from the right hits the obstacle and is partially transmitted through it as $T<1$, and part of it is also reflected creating a standing wave.

\begin{figure}[!t]
\centering
\includegraphics[width=\linewidth]{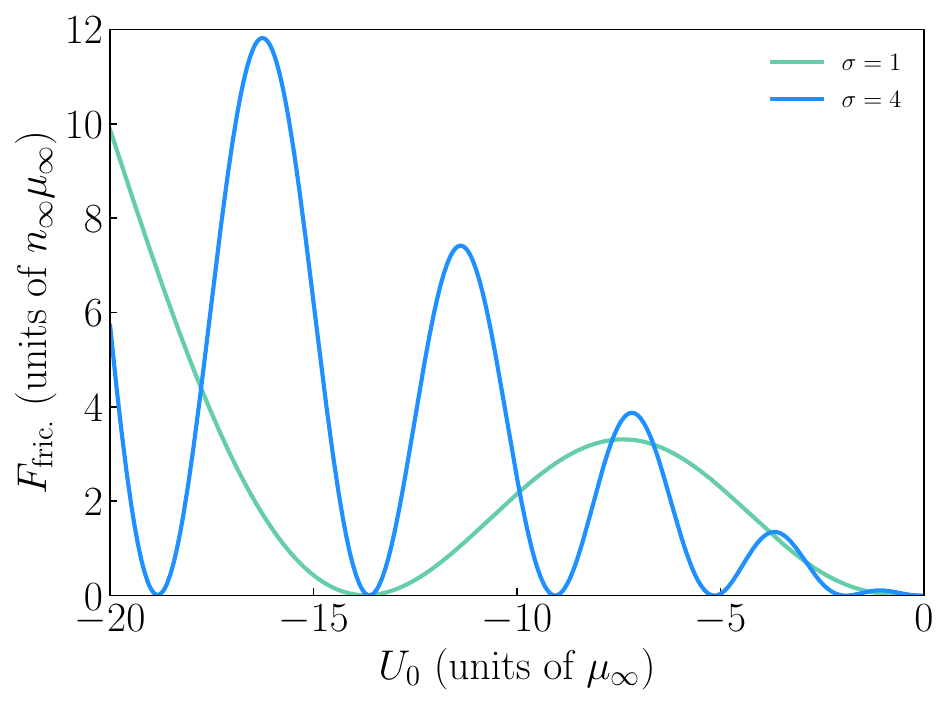}
\caption{Friction force exerted by a fluid of velocity $v_\infty=3.5$ on a square obstacle as a function of its amplitude $U_0$, for two different widths $\sigma=1$ and $\sigma=4$. The values of $U_0$ for which $F_\mathrm{fric.}=0$ are located on the lines of total transmission.}
\label{fig:force}
\end{figure}

Finally, we were able to characterise the friction exerted on the obstacle by evaluating the force
\begin{equation}
F_{\mathrm{fric.}}=-\int dx\,n(x)\partial_x U(x).
\end{equation}
We recover, as shown in fig. \ref{fig:force}, that for specific values of $(U_0,v_\infty)$ leading to $T=1$, the force drops to zero: The fluid experiences no friction along the lines of perfect transmission.

